\documentclass[prl,twocolumn,showpacs,preprintnumbers,amsmath,amssymb]{revtex4}
\usepackage{graphicx}% Include figure files
\usepackage{dcolumn}% Align table columns on decimal point
\usepackage{bm}% bold math

\def\lsim{\mathrel{\lower2.5pt\vbox{\lineskip=0pt\baselineskip=0pt
\hbox{$<$}\hbox{$\sim$}}}}
\def\gsim{\mathrel{\lower2.5pt\vbox{\lineskip=0pt\baselineskip=0pt
\hbox{$>$}\hbox{$\sim$}}}}

\newcommand{\ima}{{\mbox{Im}\,}}
\newcommand{\rea}{{\mbox{Re}\,}}
\newcommand{\be}{\begin{equation}}
\newcommand{\ee}{\end{equation}}

\newcommand{\NP}[1]{Nucl.\ Phys.\ {#1}}
\newcommand{\ZP}[1]{Z.\ Phys.\ {#1}}

\newcommand{\PL}[1]{Phys.\ Lett.\ {#1}}

\newcommand{\PR}[1]{Phys.\ Rev.\ {#1}}
\newcommand{\PRL}[1]{Phys.\ Rev.\ Lett.\ {#1}}

\begin{document}

%\preprint{hep ph/0203134}

\title{On the nature of light scalar mesons from their
large $N_c$ behavior}

\author{J. R. Pel\'aez }

\affiliation{Departamento de F{\'\i}sica Te{\'o}rica II,
  Universidad Complutense de Madrid, 28040   Madrid,\ \ Spain}

%\date{January 2002}% It is always \today, today,

\begin{abstract}
We show how to
obtain information about the 
states of an effective field theory
in terms of the underlying fundamental theory.
In particular we analyze the spectroscopic nature of meson resonances
from the meson-meson scattering amplitudes 
of the QCD low energy effective theory,
combined with the expansion in the large number of colors.
The vectors follow a  $\bar q q$  behavior, whereas the 
$\sigma$, $\kappa$ and $f_0(980)$  
scalars disappear for large $N_c$,
in support of a $\bar q\bar q qq$-like nature.
The $a_0$ shows a similar pattern, but the uncertainties
are large enough to accommodate both interpretations.
\end{abstract}

\pacs{12.39.Fe,11.15.Pg,12.39.Mk,13.75.Lb,14.40.Cs}
\maketitle

%\vspace{-.5cm}
%\section{Introduction}

Effective Quantum Field Theories are very useful
to deal systematically with the degrees of freedom
of systems when more fundamental theories are
not available or intractable.
The paradigmatic example is QCD, 
which is not able to describe
hadron dynamics at low energies, where it becomes non-perturbative.
In particular
the existence and nature
of the lightest scalar mesons is a longstanding
 controversial
issue that has recently received relevant 
experimental and theoretical contributions.
Concerning their existence, 
the implementation of the QCD
spontaneously broken chiral symmetry
leads to poles
in the pion and kaon scattering amplitudes,
associated to the most controversial states: the
$\sigma$ and the $\kappa$ \cite{newsigma}.
Such poles have been found in the most
recent charm meson decay experiments \cite{Aitala:2000xu}.
About their nature, most chiral descriptions of meson dynamics 
do not include quarks and gluons and are hard to relate to QCD,
and the spectroscopic nature is thus imposed from the start. 
In contrast, 
models with quarks and gluons, even those
inspired in QCD, have problems with chiral symmetry, small
meson masses, etc... Furthermore, both kind of models 
are usually incompatible with the chiral expansion imposed
by the low energy effective theory of QCD, known as 
Chiral Perturbation Theory (ChPT).

ChPT \cite{chpt1}
is  the most general derivative expansion of a Lagrangian,
respecting the QCD symmetries,
containing only $\pi, K$ and $\eta$ mesons. These particles 
are the Goldstone bosons of the spontaneous
chiral symmetry breaking of massless QCD and 
are the QCD low energy degrees of freedom.
For two-meson scattering  ChPT is an expansion in even
powers of momenta, generically denoted as $O(p^2), O(p^4)$...,
 over a scale $\Lambda_\chi\sim4\pi f_0\simeq 1\,$GeV.
Since $u$, $d$ and $s$ quark
masses are small compared with $\Lambda_\chi$ they
are introduced as perturbations, giving rise to  
$\pi, K$ and $\eta$ masses, counted as $O(p^2)$. 
At each order in $p^2$ ChPT is the sum of all
terms compatible with the symmetries,
each multiplied by a ``chiral'' parameter, {\it thus
avoiding any bias} in setting up a chiral model of mesons.
Thus, ChPT allows for finite
Quantum Field Theory calculations, by absorbing
loop divergences order by order 
in the chiral parameters. Once the set of parameters
up to a given order is determined from experiment, it describes, 
to that order, any other process involving mesons. 
At leading order there is only one parameter, the pion
decay constant in the chiral limit, $f_0$, 
that fixes $\Lambda_\chi$, so that
all underlying theories breaking chiral symmetry at the same scale
have the same leading term.
Different underlying dynamics manifest through different 
chiral parameters at higher orders.
We show in Table I the $L_i$ parameters that determine
meson-meson scattering up to $O(p^4)$.
As usual after renormalization, they
depend on an arbitrary 
 regularization scale $\mu$:
\begin{equation}
 L_i(\mu_2)=L_i(\mu_1)+\frac{\Gamma_i}{16\pi^2}\log\frac{\mu_1}{\mu_2}.
\end{equation}
where $\Gamma_i$ are constants given in \cite{chpt1}.
Of course, in physical observables the $\mu$ dependence is canceled
through the regularization of the loop integrals. 

The large $N_c$ expansion \cite{'tHooft:1973jz}
is the only
analytic approximation to QCD in the whole
energy region. Remarkably, it 
provides a clear definition of $\bar qq$ states that become
bound states when $N_c\rightarrow\infty$.
ChPT being the low energy QCD effective theory,
the $N_c$ scaling
of its $L_i$ parameters, listed in Table I, has been obtained in
\cite{chpt1,chptlargen}. 
In addition, the $\pi,K,\eta$ masses scale as $O(1)$ and 
$f_0$ as $O(\sqrt{N_c})$.
There is still the question of what is the renormalization scale
at which the $N_c$ scaling should be applied to the $L_i(\mu)$.
The scale dependence is certainly
suppressed by $1/N_c$ for $L_i=L_2,L_3,L_5,L_8$, but not for
$2L_1-L_2, L_4,L_6$ and $L_7$.
Even though the subleading pieces will become proportionally
less important at large $N_c$, the logarithmic terms 
can be rather large for $N_c=3$ \cite{Pich:2002xy}. 
The separation between the large $N_c$ leading and
subleading parts of the {\it measured} $L_i$ 
is not possible, but the leading $N_c$ estimates 
work well around $\mu\simeq\Lambda_\chi\simeq 1\,$GeV
(as we will check below with the vector mesons).
Indeed, the $\mu$ 
where the $N_c$ scaling applies has been estimated 
between 0.5 and 1 GeV \cite{chpt1}.

\begin{table}[hbpt]
\begin{tabular}{|c|c|c|c|}
\hline
Parameter & ChPT \cite{chpt1,BijnensGasser}&
IAM&  
%Large $N_c$&
Large $N_c$ 
\\
$\times 10^{-3}$&$\mu=770$ MeV&$\mu=770$ MeV&
%$N_c=3$&
behavior
\\
\hline
$2 L_1- L_2$
& $-0.55\pm0.7$
& $0.0\pm0.1$ 
%& $0$
& $O(1)$
\\
$L_2$
& $1.35\pm0.3$ 
& $1.18\pm0.10$ 
%& $1.8$
& $O(N_c)$\\
$L_3 $  &
 $-3.5\pm1.1$&
$-2.93\pm0.14$ 
%&$-4.3$
& $O(N_c)$
\\
$L_4$
& $-0.3\pm0.5$& 
$0.2\pm0.004$ 
%& $0$
& $O(1)$\\
$L_5$
& $1.4\pm0.5$
& $1.8\pm0.08$ 
%& $2.1$
& $O(N_c)$
\\
$L_6$
& $-0.2\pm0.3$
& $0.0\pm0.5$ 
%&$0$
&$O(1)$\\
$L_7 $  &
$-0.4\pm0.2$&
$-0.12\pm0.16$ &
%$-0.3$&
$O(1)$
\\
$L_8$
& $0.9\pm0.3$& $0.78\pm0.7$ 
%&$0.8$
&$O(N_c)$\\
\hline
\end{tabular}
\caption{Chiral parameters
from ChPT and the unitarized amplitudes (IAM) and
their leading $N_c$ scaling from QCD. 
} 
\label{elesln}
\end{table}

Since ChPT is an expansion in momenta and masses,
it is limited to low energies. 
As the energy grows, the ChPT truncated series
will violate unitarity.
Nevertheless, in recent
years ChPT has been extended to higher energies by means of unitarization 
\cite{Dobado:1996ps,Oller:1997ng,Guerrero:1998ei,GomezNicola:2001as,Oller:1997ti}. 
The main idea is that when projected into partial waves of definite
angular momentum $J$ and isospin $I$, physical amplitudes $t$ should satisfy
an elastic unitarity condition:
\begin{equation}
  \ima t =\sigma \vert t\vert^2 \Rightarrow \ima \frac{1}{t}=-\sigma \Rightarrow
t=\frac{1}{\rea t^{-1} - i \sigma},
\end{equation}
where $\sigma$ is the phase space of the two mesons, a well known function.
A ChPT calculation up to a given order 
does not satisfy this constraint,
since the powers of momenta will not match in the left hand equality.
However, from the right hand side we note that to have a unitary
amplitude we only need $\rea t^{-1}$, and for that 
we can use the ChPT expansion; this is the
Inverse Amplitude Method (IAM) \cite{Dobado:1996ps}.  The results of this simple resummation
are remarkable, since it generates resonances 
not initially present in ChPT like
the $\rho$, $K^ *$, the $\sigma$ and the $\kappa$, ensuring unitarity
in the elastic region and respecting the low energy ChPT expansion.
When inelastic two-meson processes are present 
all partial waves $t$ between
all physically accessible states can be gathered in a symmetric
$T$  matrix.
Then, the IAM generalizes 
to $T\simeq(\rea T^{-1}-i \Sigma)^{-1}$ where $\Sigma$ is a diagonal
matrix containing the phase spaces of all accessible two meson states,
again well known 
\cite{Oller:1997ng,Guerrero:1998ei,GomezNicola:2001as,Oller:1997ti}. 
With this generalization it was recently
shown  \cite{GomezNicola:2001as} that, using the one-loop ChPT calculations,
it is possible to generate the four resonances mentioned above
together with the $a_0(980)$, the 
$f_0(980)$ as well as the octet $\phi$, extending
the ChPT description of two body $\pi$, K or $\eta$
scattering up to 1.2 GeV, but keeping simultaneously
the correct low energy expansion and with chiral parameters
compatible with standard ChPT. 
We show in Table I the $L_i$ obtained
from a recent update of an IAM fit to the scattering data \cite{GomezNicola:2001as}.

One may wonder how robust are these results.
Similar unitarization methods \cite{Oller:1997ti,Nieves:1999bx} 
lead to similar results. In particular the $\sigma$ and $\kappa$
are obtained as soon as one 
requires chiral symmetry and unitarity. The use of ChPT ensures
that we are not forgetting any contribution up to $O(p^ 4)$,
and that we could extend it to higher orders if we wished.
Indeed, the IAM has been applied to $\pi\pi$ up to $O(p^ 6)$
finding basically the same results \cite{Nieves:2001de}. 
Also, with an order of magnitude
estimate for the leading $O(p^ 8)$ contribution it is even possible
to go up to 1400 MeV in the $J=2$ channel, generating the $f_2(1250)$
\cite{Dobado:2001rv}.
One could worry about crossing symmetry, but
it has been shown that the amount of crossing violation is 
smaller than the present experimental uncertainties \cite{Nieves:2001de}.
Furthermore, using the Roy equation formalism for $\pi\pi$, which respects also
crossing symmetry, it has been recently found \cite{Ananthanarayan:2000ht} 
a similar pole for the $\sigma$.
Unlike the IAM, all these improvements in $\pi\pi$ have not been applied to 
other processes because they become much more complicated.

Since we are interested in the specific underlying QCD dynamics,
we have to consider at least $O(p^ 4)$
terms. It is possible to describe 
the scalar channels with the leading order plus 
a cutoff (or another regularization parameter)
playing the role of some combination of higher order parameters. 
However, for the vector resonances we need at least the $O(p^ 4)$ parameters,
and we want to generate the vectors to test that our approach
is able to identify first of all the well established $\bar qq$ states,
and the scale $\mu$ where the $N_c$ scaling applies.
For those reasons, we use the one-loop $O(p^ 4)$ 
meson-meson
 scattering amplitudes unitarized with the IAM \cite{GomezNicola:2001as}.
Different IAM fits are due to different ChPT truncation schemes
equivalent up to $O(p^ 4)$ and to the estimates of 
the large systematic uncertainties 
in the data; we have chosen a representative fit in Table I, 
but the results are similar for other sets.
Note that these ChPT amplitudes are fully
renormalized in the $\overline{MS}-1$ scheme,
and therefore scale independent. Hence
all the QCD $N_c$ dependence appears correctly
through the $L_i$ 
and cannot  hide in any spurious parameter.
If we had kept just the leading order and a regularization scale
or a cutoff, we would not know if that cutoff is playing the role of,
for instance, $L_2$ or $L_8$,
or any other $O(N_c)$ combination of $L_i$.

Let us then scale $f_0\rightarrow f_0 \sqrt{N_c/3}$ for $m=\pi,K,\eta$,
and $L_i(\mu)\rightarrow L_i(\mu)(N_c/3)$ for $i=2,3,5,8$, keeping
the masses and $2L_1-L_2,L_4,L_6$ and $L_7$ constant.
Fig.1 shows, for increasing
$N_c$, the modulus of the 
$(I,J)=(1,1)$ and $(1/2,1)$ amplitudes. We see the
Breit-Wigner shape of the $\rho$
and $K^*(892)$ vector resonances, respectively, becoming
narrower as $N_c$ increases, but with a peak 
at an almost constant position. 
In contrast all over both the $\sigma$ and $\kappa$ regions
the amplitudes decrease with larger $N_c$.

\begin{figure}[h]
\includegraphics[scale=.92]{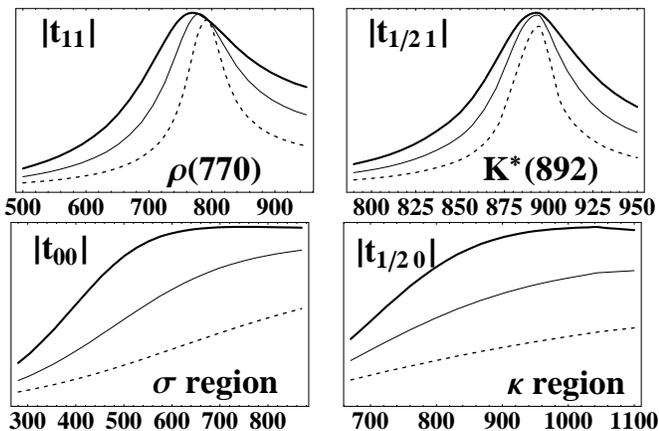}% Here is how to import EPS art
\caption{\rm Modulus of amplitudes in different meson-meson channels
for $N_c=3$ (thick line) $N_c=5$ (thin continuous line) and $N_c=10$ (thin dotted line), scaled at $\mu=770\,$MeV.
}
\end{figure}

In Fig.2 we show the evolution
of the $\rho$ and $K^*$ pole positions,
related to the mass and width as $\sqrt{s_{pole}}\simeq M-i \Gamma/2$
(as for Breit-Wigner resonances, but abusing the notation
for the rest).
We have normalized both $M$ and $\Gamma$
to their value at $N_c=3$ in order to compare
with the $\bar{q} q$ expected behavior:
$M_{N_c}/M_3$ constant and  $\Gamma_{N_c}/\Gamma_3\sim 3/N_c$.
The agreement is remarkable, not only
qualitatively, but {\em also quantitatively} within
the gray band that covers the uncertainty on the scale
$\mu=0.5-1\,$GeV where to apply the large $N_c$ scaling.
We have checked that outside that band, the behavior starts 
deviating from that of $\bar qq$ states, which confirms that
the expected scale range where the large $N_c$ scaling 
applies is correct.
In contrast, the $\sigma$ and $\kappa$ poles show a totally
different behavior, since 
{\em their width grows with} $N_c$, in
conflict with a ${\bar qq}$ interpretation. 
This was also suggested using the ChPT leading order
unitarized amplitudes with a regularization scale \cite{Oller:1997ti,Harada:2003em}.

\begin{figure}[h]
\includegraphics[scale=.81]{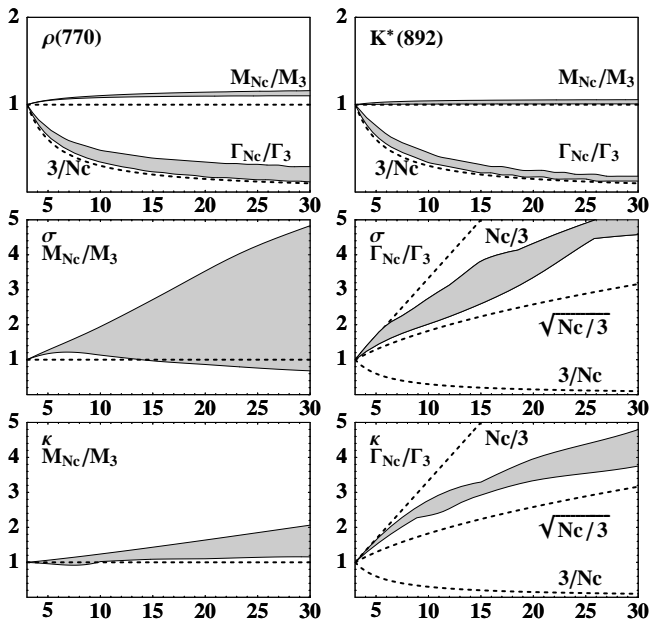}% Here is how to import EPS art
\caption{\rm $N_c$ dependence of the $\rho$, $K^*$, $\sigma$ and $\kappa$
pole positions, defined as $\sqrt{s_{pole}}\simeq M-i\Gamma/2$, normalized
to their $N_c=3$ value. The dashed lines show different $N_c$ scaling laws,
and the gray areas cover the uncertainty in $\mu\simeq 0.5-1\,$GeV.}
\end{figure}

\begin{figure}[h]
\includegraphics[scale=.7]{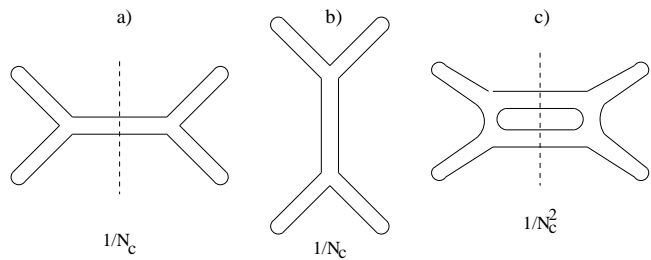}% Here is how to import EPS art
\caption{\rm Representative diagrams contributing to meson-meson scattering
and their $N_c$ scaling.}
\end{figure}

In order to determine what these states could be,
we have checked that {\em in the whole} $\sigma$ and $\kappa$ regions,
the corresponding $\ima t\sim O(1/N_c^2)$ and $\rea t\sim O(1/N_c)$.
Diagrammatically, imaginary parts can only be generated from 
graphs like those in Fig.3.a and 3.c, when the intermediate state (represented
by the dotted line) is physically accessible. But Fig.3.a has an intermediate
$\bar qq$ meson, with mass $M\sim O(1)$ and $\Gamma\sim 1/N_c$, so that
at $\sqrt{s}\simeq M$ we expect $\ima t\sim O(1)$ and a peak, as it is indeed
the case of the $\rho$ and $K^*$. Therefore, the $\sigma$ and $\kappa$ 
do not get imaginary parts from graphs like that of Fig.3.a, although
they get a $1/N_c$ contribution to the real part from Fig.3.b, 
usually interpreted as $\rho$ or $K^*$ t-channel exchange, 
respectively. 
The leading s-channel contribution in terms of quarks and gluons
comes from the graph in Fig.3.c. For the $\kappa$, 
which is a strange particle, this means a leading $\bar q \bar q qq$ (or two
meson)
contribution. This kind of states are predicted to unbound and  
become the meson-meson continuum in the $N_c\rightarrow\infty$ limit \cite{Jaffe}.
The same interpretation holds for the sigma,
but Fig.3.c also corresponds to a glueball exchange, 
that we cannot exclude with these $N_c$ arguments  alone.
However, the lightest glueball is expected with a mass higher than 1 GeV
and  $SU(3)$ symmetry would suggest that the 
$\kappa$ and the $\sigma$ should be rather similar.
Thus, a dominant $\bar q \bar q qq$  component for the $\sigma$ seems the
most natural interpretation, although it can certainly have some glueball
mixing.

Finally, Fig.4 shows the large $N_c$ behavior
in the $f_0(980)$ and $a_0(980)$ region,
which are more complicated due to the distorsions caused by the nearby $\bar KK$ threshold.
 The $f_0(980)$ is 
characterized by a sharp dip in the amplitude
that vanishes at large $N_c$, contrary
to the expectations for a $\bar qq$ state.
Note that for smaller $N_c$, the position of the 
disappearing dip changes but
for $N_c>5$ it follows again the $1/N_c^2$ scaling compatible 
with $\bar q\bar q qq$ states or glueballs.
The $a_0(980)$ behavior is more complicated.
When we apply the $L_i(\mu)$ large $N_c$ scaling
at $\mu=0.55-1\,$ GeV, its
 peak disappears, suggesting that this is not a $\bar q q$
state, and the imaginary part of the amplitude
follows roughly the $1/N_c^2$ behavior
in the whole region. 
However, as shown in Fig.5, the peak does not vanish at
large $N_c$ if we take $\mu=0.5\,$GeV. Thus we cannot rule
out a possible $\bar q q$ nature, or a sizable mixing,
although it shows up in
an extreme corner of our uncertainty band. 
For other recent large $N_c$ arguments in a chiral context
see \cite{Cirigliano:2003yq}

\begin{figure}[h]
\includegraphics[scale=.5]{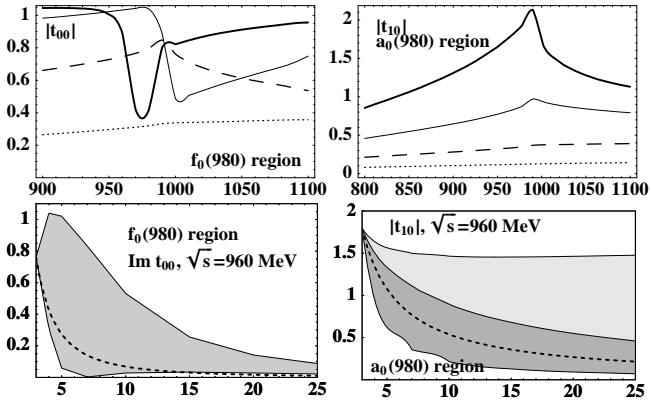}% Here is how to import EPS art
\caption{\rm Top: Modulus of $(I,J)=(0,0),(1,0)$ amplitudes 
for $N_c=3$ (thick line) $N_c=5$ 
(thin continuous line), $N_c=10$ (dashed)  and $N_c=25$
 (thin dotted line), scaled at $\mu=770\,$MeV. Bottom:
Imaginary part and modulus of amplitudes versus $N_c$ in the resonant regions.
Dark gray areas cover $\mu=0.55-1\,$ GeV, the light gray area 
covers the uncertainty down to $0.5\,$GeV.}
\end{figure}

\begin{figure}[h]
\includegraphics[scale=.6]{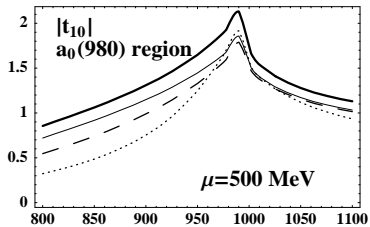}% Here is how to import EPS art
\caption{\rm Modulus of $(I,J)=(1,0)$ amplitude
for $N_c=3$ (thick line) $N_c=5$ 
(thin continuous line), $N_c=10$ (dashed)  and $N_c=25$
 (thin dotted line), scaled at $\mu=500\,$MeV. }
\end{figure}

In conclusion, we have shown how by changing 
effective Lagrangian parameters according to some specific rules
dictated by the underlying dynamics, we can learn
about the structure of the states at the fundamental level.
In particular, we have shown that the QCD large $N_c$ scaling
of the unitarized meson-meson amplitudes
of  Chiral Perturbation Theory 
is in conflict with a $\bar qq$ nature for the lightest scalars
(not so conclusively for the $a_0(980)$), 
and strongly suggests a 
$\bar q \bar qqq$ or two meson main component, maybe with some mixing
with glueballs, when possible. 

The techniques here presented could be easily applied in
other frameworks were unitarized effective Lagrangian
amplitudes already exist, as Heavy Baryon Chiral 
Perturbation Theory \cite{GomezNicola:2000wk}
or the strongly interacting symmetry breaking sector 
of the Standard Model \cite{Dobado:1989gr}.
With somewhat more effort they could also be applied
when the fundamental theory is intractable
but has a simpler description in terms of effective Lagrangians.

{\bf Note added:} The idea of this work
and the pole movements
were presented by the author in two workshops \cite{Pelaez:2003ip}.
While completing the calculations and the manuscript
the results without the scale uncertainties
have been confirmed \cite{Uehara:2003ax}  for all resonances, using the
approximated IAM \cite{Oller:1997ng}.

%I wish to thank A. Andrianov, D. Espriu, A. G\'omez Nicola, F. Kleefeld,
%R. Jaffe, E. Oset, J. Soto and  M. Uehara for their comments and suggestions.
%Work supported by 
%the Spanish CICYT projects,
%BFM2000-1326 and BFM2002-01003 and the 
%E.U. EURIDICE network contract no. HPRN-CT-2002-00311.

\bibliography{apssamp}% Produces the bibliography via BibTeX.

%\vspace{ .6cm}

\end{document}